# INFORMLEDGE SYSTEM

## *A Modified Knowledge Network with Autonomous Nodes using Multi-lateral Links*

Dr T.R. Gopalakrishnan Nair
*Director-RIIC, Dayananda Sagar Institution, 4th Floor, New Business Block, KumaraswamyLayout, Bangalore, India*
*trgnair@yahoo.com*

Meenakshi Malhotra
*Associate Member-RIIC, D. S. Institution, 4th Floor, New Business Block, KumaraswamyLayout, Bangalore, India*
*Uppal_meenakshi@yahoo.co.in*

Abstract: Research in the field of Artificial Intelligence is continually progressing to simulate the human knowledge into automated intelligent knowledge base, which can encode and retrieve knowledge efficiently along with the capability of being is consistent and scalable at all times. However, there is no system at hand that can match the diversified abilities of human knowledge base. In this position paper, we put forward a theoretical model of a different system that intends to integrate pieces of knowledge, *Informledge System* (ILS). ILS would encode the knowledge, by virtue of knowledge units linked across diversified domains. The proposed ILS comprises of autonomous knowledge units termed as Knowledge Network Node (KNN), which would help in efficient cross-linking of knowledge units to encode fresh knowledge. These links are reasoned and inferred by the Parser and Link Manager, which are part of KNN.

## 1 INTRODUCTION

Enormous amount of demand exists in Artificial Intelligence and cognitive systems to facilitate knowledge storage and retrieval with consistency and scalability. As from Weishan Zhang, Thomas Kunz (2006), certain amount of research has already taken place that helps in systematically connecting words to formulate sentences of admissible meanings. Though the efforts are continued in this field, no brakethrough has been reported for evolution of an auto routed knowledge component that is not only capable of storage, but can intelligently parse and link. *Informledge System* transforms information stored into meaningful knowledge by virtue of multi-lateral links which are inferred and reasoned by parser and link manager.

This paper is organized as follows. Section 2 discusses Research Background. Section 3 describes Research Objective and Section 4 gives composition of Knowledge Network Node. Section 5 gives the Conceptual Model of ILS. Section 6 shows how knowledge is encoded into ILS and other knowledge representations. Finally, we conclude in section 7.

## 2 RESEARCH BACKGROUND

Since early 1980s ontologies have been vital part for knowledge representation in various fields of AI and Semantic Web. The term ontology which originated from the field of philosophy is meant to represent what exist. In computer science theory, "ontology is formal, explicit specification of a shared conceptualization" (Thomas R. Gruber, 1993). For knowledge sharing and reuse, numerous representations have been devised to structure the stored information.

### 2.1 Knowledge Bases

The need for the systems to encode and retrieve information intelligently has lead to the creation of knowledge base systems. Knowledge comprises of concepts, theory, facts and rules which are modelled using ontologies. The necessity to integrate domain specific ontologies and reuse data elements from existing system has led towards Standard Upper Ontology (SUO). SUO provide definitions for general concepts at higher-level not including domain specific concepts but acts as a foundation for more specific domain ontologies (Ian Niles, Adam Pease, 2001). The upper ontologies available are: OpenCyc, BFO, DOLCE and DnS , GFO ,IDEAS ,SUMO , IDM , Biomedical ontology , COSMO.

However, Ontology development faces new challenges as listed below (Ontology Development Pitfalls, 2005, para 1):

- Development of domain ontologies has led to semantic heterogeneity (Robert M. Colomb, 2007, 'Upper Ontologies', para 2).
- It fails to distinguish between different relationships like 'instance-of' relationship.
- Problem in reasoning the events that are events relations between concepts
- Failure to model change in facts over time.

## 2.2 Semantic Web

The ever increasing content on World Wide Web requires it to be shared and reused, which gave way to the development of Semantic Web (Ivan Herman, 2010). According to Sean Bechhofer, Ian Horrocks and Peter F. Patel-Schneider (2003) Semantic Web require:

- Metadata to access the shared information,
- Ontologies to provide vocabulary for annotations
- Standard web ontology language to have a common syntax before we can share semantics

Semantic Web imposes new challenges as listed by Kenneth, Paulo C. G. Costa, Mieczyslaw M. Kokar, Trevor Martin, Thomas Lukasiewicz (2008):

- Pages must be semantically annotated through processes that are mostly manual and require good engineering skill.
- Generating metadata is the requirement.
- People's privacy could become compromised.
- Logical contradictions, inconsistencies, which will inevitably arise when ontologies from separate sources are combined.
- It extracts information mainly from web which is typically incomplete and uncertain.
- There is a risk of disingenuous information, as anyone can publish anything on the web.

## 3 RESEARCH OBJECTIVE

The proposed system (ILS) aims at formulating knowledge through the intelligent links. The links form the vital part of this system, which is not just an arc but is a detailed specifier for the linkage i.e. having information about type, direction, connectivity, source and destination nodes, creation time, usage, etc. of the link. It can be a connection between two nodes/ concepts/ entities/ events/ elements or even two links.

The system is knowledgeable by virtue of its understanding i.e. it can classify, correlate and extrapolate the information stored using the linkages. ILS provides knowledge-to-knowledge interaction which is not possible if we are storing merely text. This interaction is possible with multi-lateral linking. Ontologies represent concepts, but concepts change with time based on occurrences of some events. In the proposed system, any event or a change in a fact is a knowledge evolution and is handled via link creation, modification or deletion.

## 4 KNOWLEDGE NETWORK NODE

ILS encompasses set of autonomous nodes termed as Knowledge Network Node (KNN). KNN will help in automatically encoding new knowledge within the existing ILS with the help of its four quadrants (shown in Figure.1):

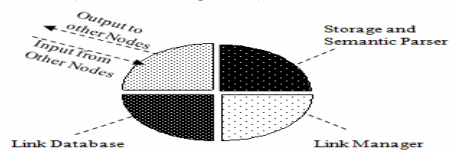

Figure 1: Knowledge Network Node

- Input / Output: This part has the knowledge to be encoded and the identified knowledge units as input and the next KNN as output.
- Storage and Semantic Parser: It stores the knowledge unit and its meaning. In addition to this it has a parser which is responsible for generation of link which is reasoned and inferenced by the parser itself.
- Link Manager: It helps in addition, deletion, classification and prioritization of the links. Prioritization of links is based on many factors such as: links connecting KNN in demand, recently used links, etc.
- Link Database: It holds all the information about the links of encoded knowledge. It stores link attributes i.e. link description, type: additive/integrative/inclusive, creation time, source, destination, last used, status, etc.

During knowledge retrieval, the existing links helps in formulation of knowledge, from the destination node, an instance of this thread can be easily pulled out as shown in Figure 2. This link information will also help in faster encoding of new knowledge, by using existing KNNs.

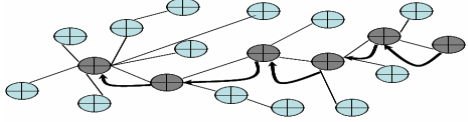

Figure 2 Extraction of knowledge thread

## 5 INFORMLEDGE SYSTEM

Informledge System (ILS) is a collection of linked KNNs. ILS targets at encoding knowledge meaningfully and with less abrasion, to make the retrieval meaningful and easy. ILS has Knowledge Library which consists of an index of all KNNs present with the system. Knowledge Library serves as an entry into ILS. Encoding can start from any arbitrary KNN termed as KNN1 and this is identified by Knowledge Encoder (KE) which is a part of ILS architecture, as shown below in Figure 3. The first knowledge unit can be considered as the head of the knowledge from where we splinter to the subsequent nodes through intelligent links.

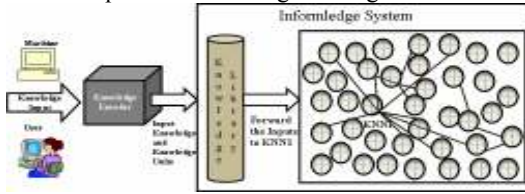

Figure 3: Conceptual Model of Informledge System

The knowledge unit extracted by the KE is fetched and knowledge to be encoded is provided as input to KNN1. Semantic Parser at KNN1 parses the knowledge and gets the links to be established between this and next KNN. Link Manager appends links to the database, thus connecting the 2 KNNs. The Parser at KNN1 creates output by appending the address of next KNN with the input, which is then taken as input to the subsequent KNN. Next KNN is then fetched from the knowledge library and the same procedure is repeated for the next KNNs also. This process goes on till the knowledge is encoded completely i.e. all the knowledge units are linked. Thus encoding process advances from one to other KNN, through the ingrained intelligence of links. Links in ILS carries information to the destination KNN that parse this link information leading to creation of links to subsequent nodes. Knowledge of ILS will evolve with increase in number of links, resulting from more and more knowledge being encoded into ILS. Also, new knowledge can be encoded using existing KNNs and freshly created links.

## 6 ENCODING

Efficient knowledge retrieval is based on how we encode knowledge and how we represent knowledge.

### 6.1 Encoding into existing KBs

Ontologies are used to represent concepts. Classes are used to represent concepts and collections, where an instance of a class represents individuals, and attributes represent individual or class properties. Along with classes, class interconnections, assertions, rules and restrictions are part of any ontology (H´ector D´ıez-Rodr´ıguez and Guillermo Morales-Luna, Jos´e Oscar Olmedo-Aguirre, 2008, '3.2' Sec, para 2). To encode something like "African Lions are Strong" Cyc Ontology will need to create terms or classes for Africa and Lion. OWL builds on RDF and RDF Schema which adds more vocabulary for describing properties and classes, relations between classes (e.g. disjointness), cardinality (e.g. "exactly one"), equality, richer typing of properties, characteristics of properties (e.g. symmetry), and enumerated classes (Ivan Herman,2007 W3C).

### 6.2 Encoding into Semantic Web

According to Sean Bechhofer, Ian Horrocks and Peter F. Patel-Schneide r(2003) Semantic Web uses URI on web as in Figure 4, to link datasets like Dbpedia, GeoNames, FOAF, VIAF,Freebase etc.

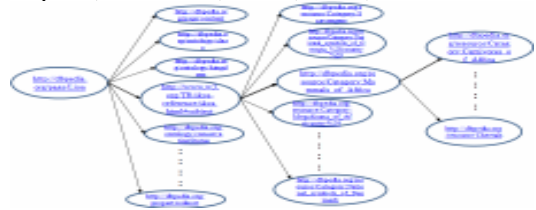

Figure 4 Data linkages in Semantic web

To encode Africa Lions are Strong, the properties of entities need to be updated along with creation of new data and thus linking it. For this we need to create semantic web page for lion and define a class for it. Then need to define properties like to which category they belong and where they belong and linking the same to the page of the sub-category of lion, carnivore of Africa. The knowledge about African Lion being strong is part of the value of abstract property of this page.

### 6.3 Encoding into ILS

To begin with, the knowledge "African Lions are Strong" is given as input to KE, which screens it and recognizes the sub-knowledge units as: Africa, Lion and Strong, represented as KNN1, KNN 2 and KNN3 as shown in Figure 5. The three KNNs belong to three different domains, but the structure used to represent the three is the same. From KNNs the links are there to n-number of other knowledge nodes, forming a cloud of knowledge.

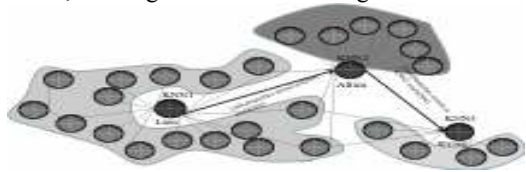

Figure 5 Encoding into ILS depicting knowledge Clouds

The link between KNN1 and KNN2 depicts that the two sub knowledge entities are connected, which means knowledge about lions in Africa and Africa having lions both are there. Moving further, if KNN2 and KNN3 are not linked i.e. the knowledge African lion being strong is not there. Then the Link manager at KNN2 would create a link and would update the link database at the respective KNN's i.e. KNN1, KNN2 and KNN3. These links implies that Africa, Lion and Strong are now linked. Thus when both the links are taken together, the following knowledge could be retrieved: "African Lions are strong", "Lions of Africa are strong" and "Strong Lions are in Africa".

Semantic web will not be able to link back to same thing i.e. Strong to African Lion, unless another entity for strong is made. And in other knowledge bases the reverse assertion need to be inserted. However, using the link properties in ILS, we extract the knowledge thread starting from KNN 'strong' and linking it to 'Lion' belonging to 'Africa'. So far knowledge is fed manually into ILS but to simulate real human knowledge, we could use Natural language processing and domain experts. It shows a scalable model with high degree of rationality.

## 7 CONCLUSIONS

Informledge System deals with linked knowledge as a whole, rather than just connected words, which could be later extracted for a purpose. ILS combines the essential units of a KB i.e. words and logic, into KNN and its multi-lateral links reaching wider scopes that are not available today. ILS works fairly well with limited number of KNNs however it is required to simulate the system with real world model and need to couple it with mammoth amount of KNNs linked across domains to handle the knowledge explosion. In addition to this, advanced studies of Tensor in vector space of multiple nodes and links is under investigation to achieve further progress in managing multi-lateral links. The future work includes analysis of link properties along with its comparison to the biological properties of neurons, which would provide more insight to the knowledge handling capability of the brain.